\documentclass[fleqn,usenatbib,letters]{mnras}

\usepackage[T1]{fontenc}
\usepackage{ae,aecompl}

\usepackage{graphicx}	
\usepackage{amsmath}	
\usepackage{amssymb}	
\usepackage{xspace}
\usepackage{bm}

\title[Confirmation of TOI1408.01]{Doppler confirmation of TESS planet candidate
TOI1408.01: grazing transit and likely eccentric orbit}

\author[G.~Galazutdinov et al.]
{G.A.~Galazutdinov$^{1,2}$\thanks{E-mail: runizag@gmail.com},
R.V.~Baluev$^{3}$\thanks{E-mail: r.baluev@spbu.ru}, G.~Valyavin$^{1}$\thanks{E-mail: gvalyavin@gmail.com}, V.~Aitov$^{1}$, D.~Gadelshin$^{1}$, \newauthor
A.~Valeev$^{1,2}$, E.~Sendzikas$^{1}$, E.~Sokov$^{4}$, G.~Mitiani$^{1}$, T.~Burlakova$^{1}$, I.~Yakunin$^{1}$, \newauthor
K.A.~Antonyuk$^{2}$, V.~Vlasyuk$^{1}$, I.~Romanyuk$^{1}$, A.~Rzaev$^{1}$, M.~Yushkin$^{1}$, \newauthor
A.~Ivanova$^{5}$, A.~Tavrov$^{5}$, O.~Korablev$^{5}$\\
$^1$Special Astrophysical Observatory, Russian Academy of Sciences, Nizhnij Arkhyz 369167, Russia\\
$^2$Crimean Astrophysical Observatory, Nauchny, Crimea 298409\\
$^3$Saint Petersburg State University, 7--9 Universitetskaya Emb., Saint Petersburg 199034, Russia\\
$^4$Central Astronomical Observatory at Pulkovo, Russian Academy of Sciences, Pulkovskoe Shosse 65, 196140 Saint-Petersburg, Russia\\
$^5$Space Research Institute, Russian Academy of Sciences, 84/32 Profsoyuznaya Str., Moscow 117997, Russia\\
}

\begin{document}

\date{Accepted 2023 ....
      Received 2023 ...;
      in original form 2023 ...}

\pagerange{\pageref{firstpage}--\pageref{lastpage}} \pubyear{2022}

\maketitle

\label{firstpage}

\begin{abstract}
We report an independent Doppler confirmation of the TESS planet candidate
orbiting an F-type main sequence star TOI-1408 located 140 pc away. We present
a set of radial velocities obtained with a high-resolution fiber-optic
spectrograph FFOREST mounted at the SAO RAS 6-m telescope (BTA-6). Our self-
consistent analysis of these Doppler data and TESS photometry suggests a
grazing transit such that the planet obscures its host star by only a portion
of the visible disc. Because of this degeneracy, the radius of TOI-1408.01
appears ill-determined with lower limit about $\sim$1 R$_{\rm Jup}$,
significantly larger than in the current TESS solution. We also derive the
planet mass of $1.69\pm0.20$~$M_{\rm Jup}$ and the orbital period
$\sim4.425$ days, thus making this object a typical hot Jupiter, but with a
significant orbital eccentricity of $0.259\pm0.026$. Our solution may suggest
the planet is likely to experience a high tidal eccentricity migration
at the stage of intense orbital rounding, or may indicate possible presence of
other unseen companions in the system, yet to be detected.
\end{abstract}


\section{Introduction}
The first planet orbiting a normal star was discovered in 1995 by \citet{Mayor1995}
and immediately became a surprise for the world astronomical community,
since the existence of a planet with a mass close to that of Jupiter
on a 4-day orbit was difficult to explain by existing models at that time.
Discovery of the first transit of a hot Jupiter around HD\,209458
\citep{Henry2000,Charbonneau2000} showed a low density of such planets,
which indicates their predominantly hydrogen-helium composition.
Since then, hundreds of hot Jupiters have been discovered, but the details
of their formation are still unclear \citep{Dawson2018}. According to
the most common view, giant planets form outside the system's icy boundary
and then migrate closer to the star \citep{2007astro.ph..1485A}.

Migration can be explained by several competing theories including tidal halting; magnetorotational
instabilities; planet - disk magnetic interactions; the Kozai mechanism of a distant perturber; planet traps
and planet - planet scattering; etc (see \cite{2019A&A...628A..42H} and references in it).

In this paper, we report the discovery of the hot Jupiter TOI-1408.01
in an elongated orbit with an eccentricity of e$\approx$0.26.
Compared to some giant planets with huge eccentricities exceeding 0.9,
such as HD\,80606b \citep{Naef2001}, TOI-1408.01's eccentricity
is quite moderate, but noticeably exceeds the orbital eccentricities
of typical hot Jupiters, which are practically zero. It is likely
that TOI-1408.01 is in the final stages of rounding its orbit. The following
sections discuss the observations, analysis, and results.

\section{Observations and data reduction}
\label{Obs}

Observations were carried out with the aid of 6-m Russian Telescope (BTA) 
from 2021 to 2023 for twenty observation nights under the program
EXPLANATION (EXoPLANet And Transient events InvestigatiON, \citet{Val1,Val2}).
Observations are currently being made with the fiber-optic
high resolution echelle spectrograph FFOREST\footnote{Fiber-Feed Optical Russian Echelle SpecTrograph}.
Detailed description of the instrument given in \citet{Val3,Val4}. Briefly, the instrument is a fiber-feed
high spectral resolution echelle spectrograph with a resolving power R ($\lambda/\Delta\lambda$) beeing in the range from
$\sim$30,000 - 35,000 to $\sim$60,000 - 70,000
with further possibility to upgrade it to higher resolution, up to R$\sim$100,000. The portable fiber optic input module is mounted in the prime focus
of the 6-m telescope of the Special Astrophysical Observatory of the Russian Academy
of Sciences (SAO RAS, Russia). Presently, the instrument has two working fiber channels
with equivalent widths of fiber cores of 0.7~$\arcsec$ and 1.4~$\arcsec$.
With these cores and binning 2$\times$2 pixels we applied during our observing runs,
the instrument provides the resolving power of R$\sim$60,000 and R$\sim$30,000 respectively.
An additional calibration channel
provides simultaneous observations of spectra from ThAr and Fabry-Perout(in development) calibration sources.
It should also be noted that optical camera of the spectrograph is still under construction reducing its efficiency for
$\sim$1 stellar magnitude \citep{Val4}. Nevertheless, the spectrograph is quite effective for planetary
observations of stars as faint as 12$^m$ 
with  CCD camera designed by the Advanced Design laboratory of SAO RAS \citep{2020gbar.conf..115A}.
The CCD camera's electronics with the e2v CCD 4k$\times$4k matrix of 15$\times$15 $\mu$m pixel size provide read-out noise of 2-3 electrons.

\begin{table}
\caption{Doppler shift measurements of TOI1408. BJD - barycentric Julian date.
Radial velocity  (RV) shift values are relative to a template spectrum.}
\label{tab_rv}
\addtolength{\tabcolsep}{-3pt}
\begin{tabular}{llll}
\hline
BJD-2450000 &  RV (m\,s$^{-1}$)                  &  BJD-2450000 &  RV (m\,$s^{-1}$) \\
\hline
9532.49283  & -5604.40  $\pm$  31.20   & 9801.34242  & -5596.47  $\pm$  37.46 \\
9536.48208  & -5685.88  $\pm$  21.00   & 9802.31601  & -5582.74  $\pm$  24.54 \\
9537.22933  & -5759.83  $\pm$  29.69   & 9834.37490  & -5734.37  $\pm$  30.49 \\
9540.53171  & -5524.08  $\pm$  35.51   & 9892.26624  & -5916.12  $\pm$  18.51 \\
9624.47869  & -5580.01  $\pm$  09.78   & 9896.23607  & -5886.04  $\pm$  17.92 \\
9653.55195  & -5993.58  $\pm$  11.80   & 9899.17260  & -5636.53  $\pm$  39.96 \\
9659.45749  & -5743.06  $\pm$  23.50   & 9920.19302  & -5879.30  $\pm$  08.55 \\
9662.57867  & -5966.79  $\pm$  66.11   & 9921.42476  & -5719.47  $\pm$  51.66 \\
9798.36145  & -5681.52  $\pm$  13.89$^*$& 9947.22310 & -5734.23  $\pm$  19.50 \\
9799.44302  & -5863.04  $\pm$  27.33   &             &                        \\
\hline
\end{tabular}\\
\small
$^*$ In-transit measurement that was removed.
\end{table}

All the spectra were processed with the aid of our own software package DECH \citep{2022AstBu..77..519G}
providing all the stages of echelle image/spectra processing, measurements and analysis.
Doppler shift measurements were performed with our own code, incorporated into the DECH package. The code based on the
cross-correlation algorithm offered by \citet{1979AJ.....84.1511T}. The instrument stability is controlled and corrected
by means of simultaneous registration of ThAr spectrum in the adjacent echelle sub-orders observed through an independent
optical path (a fiber). Currently FFOREST provides measurements of Doppler shift accuracy limit $\sim$10-20 m\,s$^{-1}$ for
stellar spectra, and $\sim$1 m\,s$^{-1}$ for ThAr spectra. The accuracy of radial velocity of some data given in Table \ref{tab_rv}
is worse due to lower signal-to-noise ratio (SNR) in the corresponding spectra.

\subsection{TESS photometry}

In the MAST (Mikulski Archive for Space
Telescopes)\footnote{\url{https://archive.stsci.edu/}} we identified $193$ TESS transit
lightcurves of TOI1408. After cutting the range of $\pm 100$~min about each midtransit
point, we had $10287$ data points in total. However, different transit lightcurves
contained different number of data points, with two clear groups of transits. In the first
group there were $<25$ data points per transit, and in the second one there were about
$80-90$ data points per transit. Only a few transits appeared between these two groups.
Since our software assigns an individual model per each transit, we selected only those
transits that had $>80$ data points. There were $96$ such transits, containing $8580$ data
points. Thus we preserved $>80\%$ of the total data for our analysis.

\subsection{Stellar parameters}

Stellar parameters where initially estimated spectroscopically, using the ionization balance method
optimized for Solar type stars \citep{2002PASJ...54..451T}. Calculations were performed with the equivalent widths of $\sim$100 - 120 neutral and $\sim$12 - 15
ionized iron lines measured in our echelle spectra.
Derived $T_{\rm eff}$, $\log g$, and $\rm [Fe/H]$ are shown in Table~\ref{tab_star}. Relatively large uncertainties of these initial values are due to moderate signal-to-noise of available spectra
and limited number of measured lines of iron.
We then refined stellar parameter with the aid of  MIST isochrones \citep{2016ApJS..222....8D} with
simultaneous estimation of the star mass $M_\star$ and radius $R_\star$. The procedure was as follows: for given [Fe/H] and standard A$_v$ = 3.1 we
got all isochrone's points belonging to the main sequence and lying in the uncertainty ranges of initial $T_{\rm eff}$, $\log g$.
There are found 53 points with surprisingly low scatter as it is seen in the refined stellar parameters given in Table~\ref{tab_star}  ($T_{\rm eff}$, $\log g$ marked with MIST).

Also, based on trilinear interpolation of tables by \citet{2017A&A...600A..30C}, we derived
the limb-darkening coefficients, also shown in the table (for the quadratic limb-darkening
law). However, we must notice that there are multiple measurements of $M_\star$, $R_\star$
in the literature, and they cover rather wide spread of values (we also give them in
Table~\ref{tab_star} for reference). They are not always consistent with each other, in
particular our $R_\star$ measurement appears significantly smaller then the one provided by
TICv8 \citep{2019AJ....158..138S}. However, these data for TOI1408 look incomplete, because
they mention zero metallicity, so they might appear not very reliable. GAIA DR3 $R_\star$
estimation is based on just the photometry, so it might appear less trustable too.

We rely our further analysis on the $M_\star$, $R_\star$ derived from the MIST and our
complete set of spectrum-derived parameters. However, determining their realistic
uncertainties is more complicated, so we must recognize that they can be improved using
spectra with higher spectral resolution and SNR.

\begin{table}
\caption{TOI1408 star data}
\addtolength{\tabcolsep}{-4pt}
\label{tab_star}
\begin{tabular}{ccccc}
\hline
        & \multicolumn{2}{c}{Ionization Balance}& \multicolumn{2}{c}{MIST}\\
        \cline{2-3}
        \cline{4-5}
$T_{\rm eff}$, $\log g$ & \multicolumn{2}{c}{$6425\pm170$~K,  $4.64\pm0.30$} & \multicolumn{2}{c}{$6306\pm32$~K, $4.35\pm0.01$} \\
$\rm [Fe/H] $ &  \multicolumn{4}{c} { $7.79\pm0.12$} \\ 
\hline
Limb darkening & \multicolumn{4}{c}{$A$ = $0.2615$ $B$ = $0.2771$}\\
\hline
\multicolumn{5}{c}{Mass \& radius}\\
                 &  &MIST-derived$^*$ &   GAIA-derived    & TICv8        \\
                 \cline{3-3}
                 \cline{4-4}
                 \cline{5-5}
      $M_\star/M_\odot$ & & $1.332\pm 0.014$ &                            &  $1.375^{+0.54}_{-0.27}$     \\
      $R_\star/R_\odot$ & & $1.276\pm 0.015$ & $1.5175^{+0.030}_{-0.031}$ &  $1.485^{+0.12}_{-0.11}$  \\
\hline
\end{tabular}
\\
\small
$^*$ Pearson correlation coefficient = 0.87.\\
\end{table}

\section{Analysis and results}
\label{sec_analysis}

\subsection{Preliminary}
In our preliminary analysis of SAO radial velocities, we revealed an undoubtful periodic
signature at the planet orbital period proposed by the TESS team. However, two main issues
appeared. First, the observed RV semiamplitude infers the planet minimum mass of $\sim
1.5-2$~$M_{\rm Jup}$, and together with the TESS-provided planet radius of $r \simeq
0.700\pm 0.055$~$R_{\rm
Jup}$\footnote{\url{https://exofop.ipac.caltech.edu/tess/view_toi.php}} this would imply an
unrealistic planet density of $\rho \sim 4-6$~$\rho_{\rm Jup}$ (or $5-8$~g\,cm$^{-3}$, like in
a rocky planet). Secondly, there were clear hints of an ecentric orbit (moderately
nonsinusoidal RV variation) that needed a detailed treatment.

Large estimation of the planet density $\rho$ might indicate either an overvalued planet
mass, or an undervalued planet radius. Wrong mass estimation is unlikely, at least not by a
factor of a few. But the radius is more suspicious. According to the available TESS
solution, TOI1408.01 demonstrates a grazing transit with near-unit impact parameter. Such a
configuration implies model degeneracies that could lead to various biases in the fitted
parameters, as well as to overly optimistic uncertainties. Therefore, our first task was to
verify the planet radius estimation by performing a self-consistent Doppler+photometry fit.

\subsection{Models}
We include in our analysis the SAO Doppler time series (with one in-transit point removed)
and $96$ TESS transits described in Sect.~\ref{Obs}.

The RV model included Keplerian planetary variation (with free eccentricity) and possible
linear trend (radial acceleration).

Regarding the TESS photometric data, they come in the raw (unwhitened) and whitened
version. We used only the unwhitened one, partly because the whitened data should be fit
with a more sophisticated transit model transformed by the whitener in the same way, and
partly because the likely-biased TESS planet radius might indicate some inaccuracies of the
whitener. Unwhitened flux may involve an increased fraction of longer-period variations
appearing on the transit timescale as trends, so to handle them we added a cubic polynomial
to the model of each transit.

The remaining noise (both in the RV and the photometry) was treated as white with a
fittable jitter parameter \citep{2009MNRAS.393..969B}. We did not use more complicated noise models
like the red noise, because the number of data per a single transit (and in the RV curve)
appeared too small for that.

The whole analysis was performed using the {\sc PlanetPack} software
\citep{2013A&C.....2...18B, 2018A&C....25..221B}, which is based, eventually, on optimizing the likelihood
function of the task. The star mass and radius estimation from Table~\ref{tab_star} were
incorporated in this analysis in the form of a 2D Gaussian penalty added to the likelihood
function (so they basically served as additional input data with 2D uncertainty).

\subsection{Results}
\begin{figure*}
\includegraphics[width=0.49\linewidth]{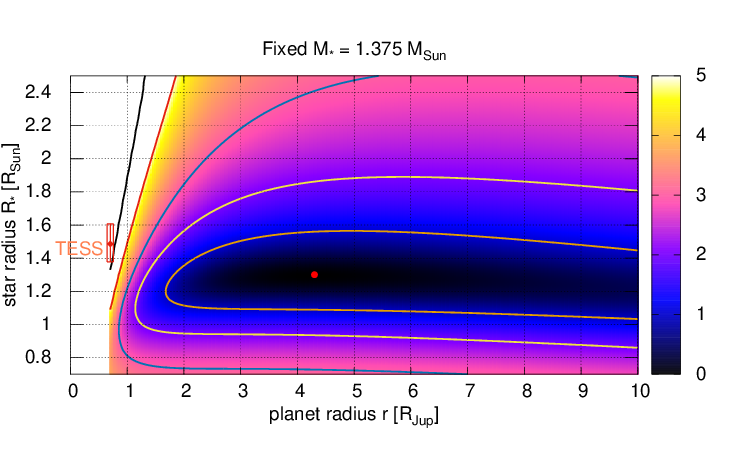}
\includegraphics[width=0.49\linewidth]{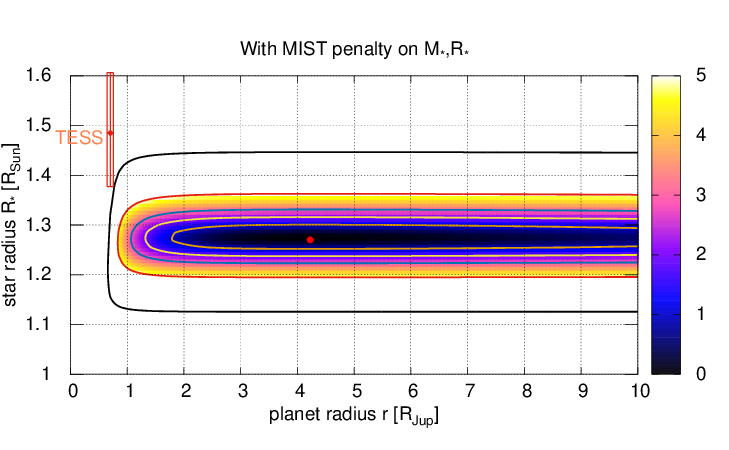}
\caption{Confidence regions for $(r, R_\star)$ constructed as level curves of the
likelihood function. Five colored curves shown in each panel outline significance levels of
$1$-sigma, $2$-sigma, $3$-sigma, $5$-sigma, and $10$-sigma. The colormap shows the same
significance (from $0$ to $5$ sigma). The left panel is for $M_\star$ fixed at the value
from TICv8, but with a priori unconstrained $R_\star$, while the right panel involves 2D
Gaussian penalty on $M_\star$ and $R_\star$ derived in this work from the MIST evolutionary
tracks.}
\label{fig_rRst}
\end{figure*}

Contrary to the nominal TESS solution, in our analysis $r$ always appeared ill-fitted.
Basically, we can arbitrarily increase $r$, simultaneously increasing the impact parameter,
so that the transit curve remains nearly unchanged. This is illustrated in
Fig.~\ref{fig_rRst}, where we plot the confidence regions in the $(r,R_\star)$ plane (based
on level contours of the likelihood function). The formal best fit corresponds to $r\sim
4$~$R_{\rm Jup}$, but even $r=\infty$ cannot be formally rejected. Nevertheless,
based on the $1-3$-sigma domains, we can put a rough lower limit on $r$ about
$1-2$~$R_{Jup}$ respectively.

Normally, transit fits should constrain the star density $\rho_\star = M_\star
R_\star^{-3}$, even without any additional astrophysical constraint on $M_\star$,
$R_\star$. But in our case, without the MIST penalty (left panel of Fig.~\ref{fig_rRst}),
all these parameters remain highly uncertain. Such additional degeneracy comes from the
fittable orbital eccentricity. Large eccentricity may affect (i) the transit duration
through the planet in-transit orbital speed and (ii) position of the transit on the RV
curve. The first effect can also mimic a change in $R_\star$, thus making it more uncertain
than usually.

Regardless of all these uncertainties, the TESS solution for $(r,R_\star)$ always lies out
of the reasonable significance domain. That is, our analysis does not confirm this
solution. In particular, the planet radius has to be significantly larger than the TESS
estimation of $0.7$~$R_{\rm Jup}$, and this further confirms our initial guess that $r$ was
significantly undervalued by the TESS team.

\begin{table}
\caption{TOI1408: Parameters and best fit values with MIST penalty on $M_\star$, $R_\star$}
\addtolength{\tabcolsep}{-2pt}
\label{tab_fit}
\begin{tabular}{lllll}
\hline
\multicolumn{2}{c}{Planet} & & \multicolumn{2}{c}{Star}\\
$K$                                & $178\pm 22$ m/s             &                      & $M_\star$         & $1.331\pm 0.014$ $M_\odot$\\
$P$                                & $4.4247110\pm 1.6\cdot 10^{-6}$ day &               & $R_\star$         & $1.274\pm 0.015$ $R_\odot$\\
$M$                                & $1.69\pm 0.20$  $M_{\rm Jup}$&                      & $\rho_\star$      & $0.643\pm 0.018$ $\rho_\odot$\\
$a$                                & $0.05804\pm 0.00020$ AU      &                      &                   &                       \\
$e$                                & $0.259\pm 0.026$             &                      & \multicolumn{2}{c}{Radial acceleration$^*$}\\
$\omega$                           & $305.2\pm 5.5$$^\circ$       &                      & $c_0$             & $-5739\pm 79$ m/s\\
$e\cos\omega$                      & $0.150\pm 0.033$             &                      & $c_1$             & $-25\pm 37$ m/s/yr\\
$e\sin\omega$                      & $-0.212\pm 0.013$            &                      &                   &                       \\
$l$                                & $273.7\pm 4.6^*$$^\circ$     &                      &       \multicolumn{2}{c}{RV jitter}\\
$i$                                & $84.801\pm 0.094$$^\circ$    &                      & $\sigma_{\rm RV}$ & $54\pm 11$ m/s\\
$r/R_\star$                        & $0.1183\pm 0.0014$           &                      &                   & \\
$r$                                & $1.5$ (fixed)$R_{\rm Jup}$   &                      &                   & \\
\hline
\end{tabular}\\
\small
$^*$Epoch JD2459000
\end{table}

\begin{figure*}
\includegraphics[width=0.49\linewidth]{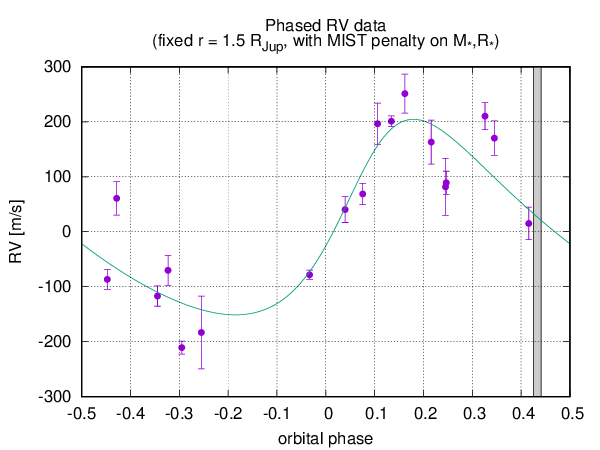}
\includegraphics[width=0.49\linewidth]{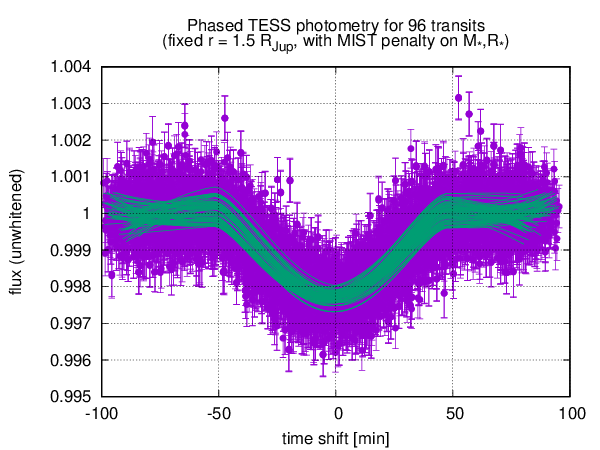}\\
\caption{Doppler (left panel) and photometric (right panel) data for TOI1408 and their best
fitting models. In the left panel we also mark the position of the transit (as a vertical
narrow band). In the right panel, each of $96$ transits has an individual polynomial trend,
so we plot a bunch of close curves that differ due to this trend.}
\label{fig_data}
\end{figure*}

Data do not constrain planet radius from the upper side, but it is obvious that $r$ cannot
be arbitrarily large, because we deal with a planetary mass object. So to construct at
least a reference (not ill-fitted) model, we have to fix $r$ at some a priori reasonable
value. For example, the fit for $r=1.5$~$R_{\rm Jup}$ is shown in Table~\ref{tab_fit}. It
corresponds to the planet density $\rho = 0.50$~$\rho_{\rm Jup}$, a priori rather realistic
value for a hot Jupiter. Smaller $r$ would imply larger density, keeping it admissible down
to $r\simeq 1$~$R_{\rm Jup}$, while larger $r$ would imply a less likely density below $0.50
\rho_{\rm Jup}$. Unfortunately, we cannot say anything more definite about $r$, regardless
of so large amount of photometric data. Available data and this reference model are shown
together in Fig.~\ref{fig_data}.

\begin{figure*}
\includegraphics[width=0.49\linewidth]{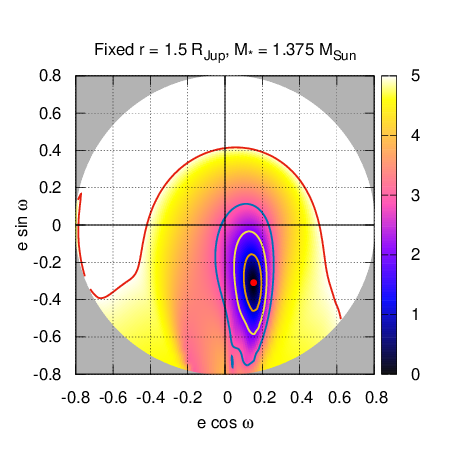}
\includegraphics[width=0.49\linewidth]{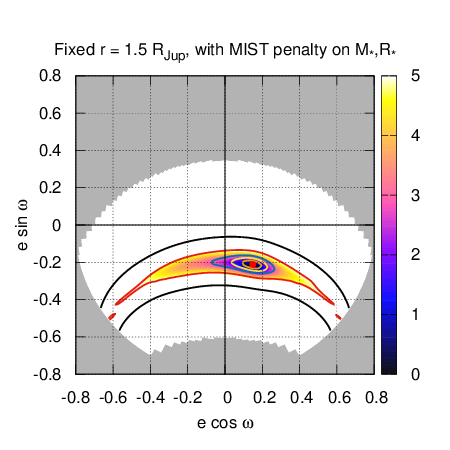}
\caption{Confidence regions for $(e\cos\omega, e\sin\omega)$. Same notes as in Fig.~\ref{fig_rRst}.}
\label{fig_ecc}
\end{figure*}

As we already noticed, Doppler data reveal important hints of an orbital eccentricity.
This is clearly seen from the shape of the RV curve (Fig.~\ref{fig_data}), and the
eccentricity estimation $e$ from Table~\ref{tab_fit} demonstrate very high formal
significance. This eccentricity estimation reveals only a weak dependence on the adopted
$r$ (with negative correlation between $r$ and $e$). However, it depends a lot on the
adopted $M_\star$, $R_\star$ penalty. If this penalty is removed, $e$ becomes much more
uncertain. This is illustrated in Fig.~\ref{fig_ecc}, where we show the 2D confidence plots
for the parameters $(e\cos\omega, e\sin\omega)$.

As we noticed above, the MIST penalty may infer overly optimistic uncertainties, so the
actuality should likely be somewhere between the left and right panels of
Fig.~\ref{fig_ecc}. Still, the eccentricity remains significant ($\sim2.5\sigma$) even in
the worst case of no-penalty. Therefore, we believe it is very likely that $e$ is nonzero,
although its particular value still need to be refined by more Doppler data. In addition to
further Doppler monitoring, it is also necessary to seek for more accurate spectrum (and
its parameters of Table~\ref{tab_star}) in order to refine astrophysical constraint on
$M_\star$ and $R_\star$.

Periodogram did not reveal any extra periodicity after subtracting the planetary signal
from the RV data, so presently there is no Doppler hints of more planets in the system.

\section{Discussion}
\label{sec_discuss}
In this paper, we explored the transit planetary candidate around the star TOI-1408, discovered by the TESS telescope. Since the period of radial velocity
oscillations coincides with the time interval between the centers of transits, we unambiguously confirm its planetary nature.
The light curve of the transits has a pronounced V-shape. This indicates the
sliding nature of the planetary disk from our spatial perspective, so we cannot estimate
the planet's radius with high accuracy. However, the mass of TOI-1408.01 measured by us, which is $\sim$1.7 M$_{Jup}$, indicates a massive gas giant. Such planets are usually
comparable to Jupiter in radius \citep{2012A&A...547A.112M} , or even larger, if we take into account the amount of insolation it receives.
According to Gaia EDR3 \citep{2021A&A...649A...1G}, the distance to TOI-1408 is 139.6 $\pm$ 0.2 pc, and based on its brightness (G = 9.195$^m$ ), the calculated luminosity of the star exceeds that of the
Sun by a factor of 3.26. Assuming zero albedo and effective heat transfer to the night side, we estimate the equilibrium temperature of the planet as $\sim$1550 $K$.
The rounding time of the orbit strongly depends on the radius of the planet and the Q$_P$ parameter \citep{2006ApJ...649.1004A}. Assuming Q$_P$ = 10$^6$ and the planetary
radius in the range of 1.3 - 1.5 R$_J$, typical for hot Jupiters with high insolation, this time is 0.61 - 1.25 Gyr.

\section*{Acknowledgements}
We obtained the observed data on the unique scientific facility "Big Telescope Alt-azimuthal"  of SAO RAS as well as made data processing with the financial support of grant No 075-15-2022-262 (13.MNPMU.21.0003)
of the Ministry of Science and Higher Education of the Russian Federation.

\section*{Data availability}
The data underlying this article are available in the article and in its online
supplementary material.

\end{document}